# Electron Scattering in Thin GaAs Quantum Wires

D.V. Pozdnyakov[1], V.O. Galenchik, A.V. Borzdov

Radiophysics and Electronics Department of Belarus State University,
Nezavisimosty av.4, 220030 Minsk, Belarus

[1]*pozdnyakov@bsu.by*

**Abstract**

In this paper the scattering rates of electrons in thin free standing GaAs quantum wires in the electric quantum limit are calculated self-consistently taking into account the collisional broadening caused by scattering processes. The following mechanisms of scattering are considered: one-dimensional acoustic and polar optical phonons, surface roughness and ionized impurities. The non-parabolicity of electron energy spectrum is also taken into account.

The interest to the simulation of charge carrier transport in structures with one-dimensional electron gas is caused by the dramatic progress in nanoelectronics technology. For the last 20–25 years the great attention has been focused on the theory of charge carrier scattering in semiconductor structures with one-dimensional electron gas [1–15]. Most of all this is concerned with the need for adequate description of scattering processes in such systems. It is known that no adequate results can be obtained from kinetic simulation without profound understanding of the scattering processes.

One of the principal spheres of nanoelectronics theory is the calculation of electron scattering rates in GaAs quantum wires [1–15]. The problems are related to the necessity of taking into account the secondary quantum effects to avoid the singularity points in the function of scattering rates versus the electron energy. The most significant effect of this kind is the collisional broadening (or electron energy uncertainty) caused by the scattering processes [4, 5, 7, 10, 14, 15]. To take it into account, the various approaches have been proposed. But, as a rule, they are either too particular or very complex from computational point of view or based on very rough assumptions (see, for example, Refs. [4, 5, 7, 10, 14, 15]). In this connection it should be noted that in Refs. [16, 17] the general and quite simple approach from the computational point of view (after some valid approximations) approach to the self-consistent, i.e. taking into account the collisional broadening, calculation of the scattering rates in the semiconductor structures with one-dimensional electron gas has been developed. The purpose of the present paper is to study the electron scattering rates for the dominant processes in thin free standing GaAs quantum wires by using the technique described in Refs. [16, 17], which was applied to the calculation of acoustic and polar optical phonon scattering rates in GaAs-in-AlAs quantum wires in Ref. [18].

Scattering rates can be calculated assuming the following valid approximations for thin free standing quantum wires [1–17, 19]: (i) the potential barrier at the boundary is infinite; (ii) phonons are one-dimensional, i.e. localized inside the wire; (iii) the electric quantum limit takes place, i.e. all the electrons are in the ground quantum state and their mean kinetic energy is much less then the value of the energy level; (iv) electrons are only scattered by the first modes of acoustic and polar optical phonons, by the surface roughness and by ionized impurities. Additionally, to simplify the numeric calculations, it can be assumed that the quantum wire has a rectangular cross-section.

Let us write down the formulae for the calculation of electron scattering rates for the dominant processes in thin free standing GaAs quantum wires taking into account the collisional broadening, non-parabolicity and the Pauli principle [1–21]

$$W_1^{f/b}(E,\Delta E) = \left(\frac{8}{3\pi}\right)^4 \left(\frac{1}{\varepsilon_\infty} - \frac{1}{\varepsilon_s}\right) \frac{e^2 \omega n \sqrt{2m_x}}{\hbar L_y L_z} \frac{(1 - f(\pm k(E+\hbar\omega)))D(E+\hbar\omega,\Delta E)}{(k(E) \mp k(E+\hbar\omega))^2 + \pi^2 (L_y^{-2} + L_z^{-2})}, \quad (1)$$

$$W_2^{f/b}(E,\Delta E) = \left(\frac{8}{3\pi}\right)^4 \left(\frac{1}{\varepsilon_\infty} - \frac{1}{\varepsilon_s}\right) \frac{e^2 \omega (n+1)\sqrt{2m_x}}{\hbar L_y L_z} \frac{(1 - f(\pm k(E-\hbar\omega)))D(E-\hbar\omega,\Delta E)}{(k(E) \mp k(E-\hbar\omega))^2 + \pi^2 (L_y^{-2} + L_z^{-2})}, \quad (2)$$

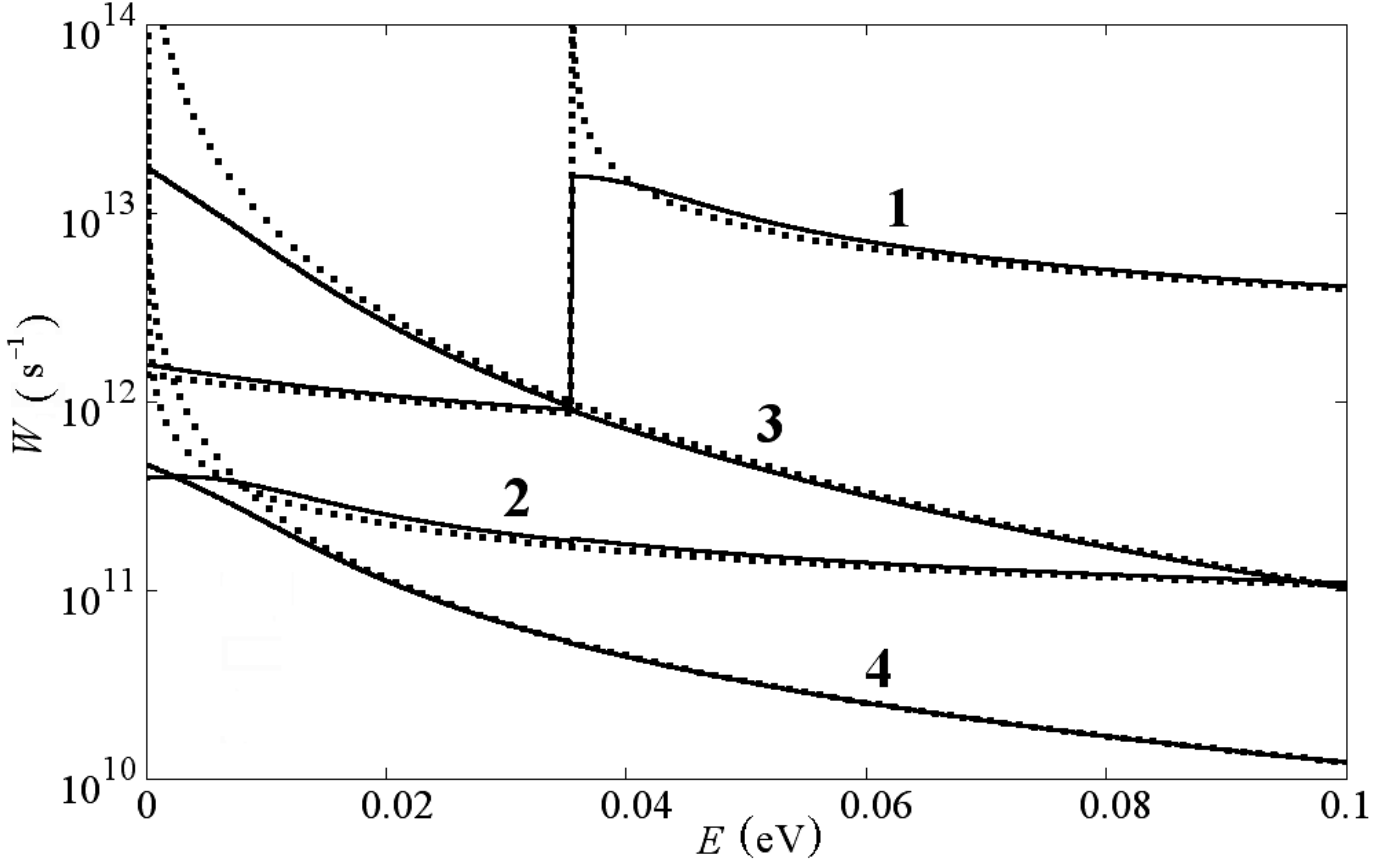

**Fig. 1.** The polar optical phonon (curves 1 – forward and backward scattering), acoustic phonon (curves 2 – only backward scattering), ionized impurity (curves 3 – only backward scattering) and surface roughness (curves 4 – only backward scattering) scattering rates of electrons in a GaAs quantum wire calculated under the following conditions:
$T = 300$ K, $L_y = L_z = 100$ Å, $\delta = 1.415$ Å, $\Lambda = 60$ Å, $N = 5\times 10^7$ $m^{-1}$, $Z = 1, f(\pm k) = 0$.
*Dot lines are the scattering rates calculated neglecting the collisional broadening.*
*Solid lines are the scattering rates calculated with the collisional broadening taken into account.*

$$W_3^{f/b}(E,\Delta E) = \left(\frac{8}{3\pi}\right)^4 \frac{2k_B T B_{ac}^2 \sqrt{2m_x}}{\hbar^2 s^2 \rho L_y L_z}(1 - f(\pm k(E)))D(E,\Delta E), \tag{3}$$

$$W_4^{f/b}(E,\Delta E) = \left(\left(\frac{\partial E_0}{\partial L_y}\right)^2 + \left(\frac{\partial E_0}{\partial L_z}\right)^2\right)\frac{\sqrt{\pi}\delta^2 \Lambda \sqrt{2m_x}}{\hbar^2}\frac{(1 - f(\pm k(E)))D(E,\Delta E)}{1 + \pi\Lambda^2 k^2((E \mp E + \Delta E)/2)}, \tag{4}$$

$$W_5^{f/b}(E,\Delta E) = F^2(k((E \mp E + \Delta E)/2))\frac{Z^2 e^4 N \sqrt{2m_x}}{2\pi^2 \varepsilon_s^2 \hbar^2}(1 - f(\pm k(E)))D(E,\Delta E), \tag{5}$$

$$D(E,\Delta E) = \Theta(E)\frac{1 + 2\eta(E + E_0)}{\sqrt{1 + \eta(E + 2E_0)}}\sqrt{\frac{\Delta E + \sqrt{E^2 + \Delta E^2}}{E^2 + \Delta E^2}}, \tag{6}$$

$$k(E) = \frac{1}{\hbar}\sqrt{2m_x E(1 + \eta(E + 2E_0))}, \tag{7}$$

$$E_0 = \frac{1}{2\eta}\left(\sqrt{1 + \frac{2\pi^2\hbar^2\eta}{m_y L_y^2} + \frac{2\pi^2\hbar^2\eta}{m_z L_z^2}} - 1\right), \tag{8}$$

$$n = \left(\exp\left(\frac{\hbar\omega}{k_B T}\right) - 1\right)^{-1}, \tag{9}$$

$$F(k) = \frac{1}{Rk} \int_0^\infty \left( RN + \frac{1}{\sqrt{1+t^2}} \right) \left( \sqrt{1+t^2} - t \right) \exp(-RNt) \sin(Rkt) dt, \quad (10)$$

$$R = \sqrt{\frac{4 L_y L_z}{\pi}}, \quad (11)$$

$$\Delta E = \frac{\hbar}{2} W_\Sigma(E, \Delta E) = \frac{\hbar}{2} \sum_{j=1}^{5} \left( W_j^f(E, \Delta E) + W_j^b(E, \Delta E) \right), \quad (12)$$

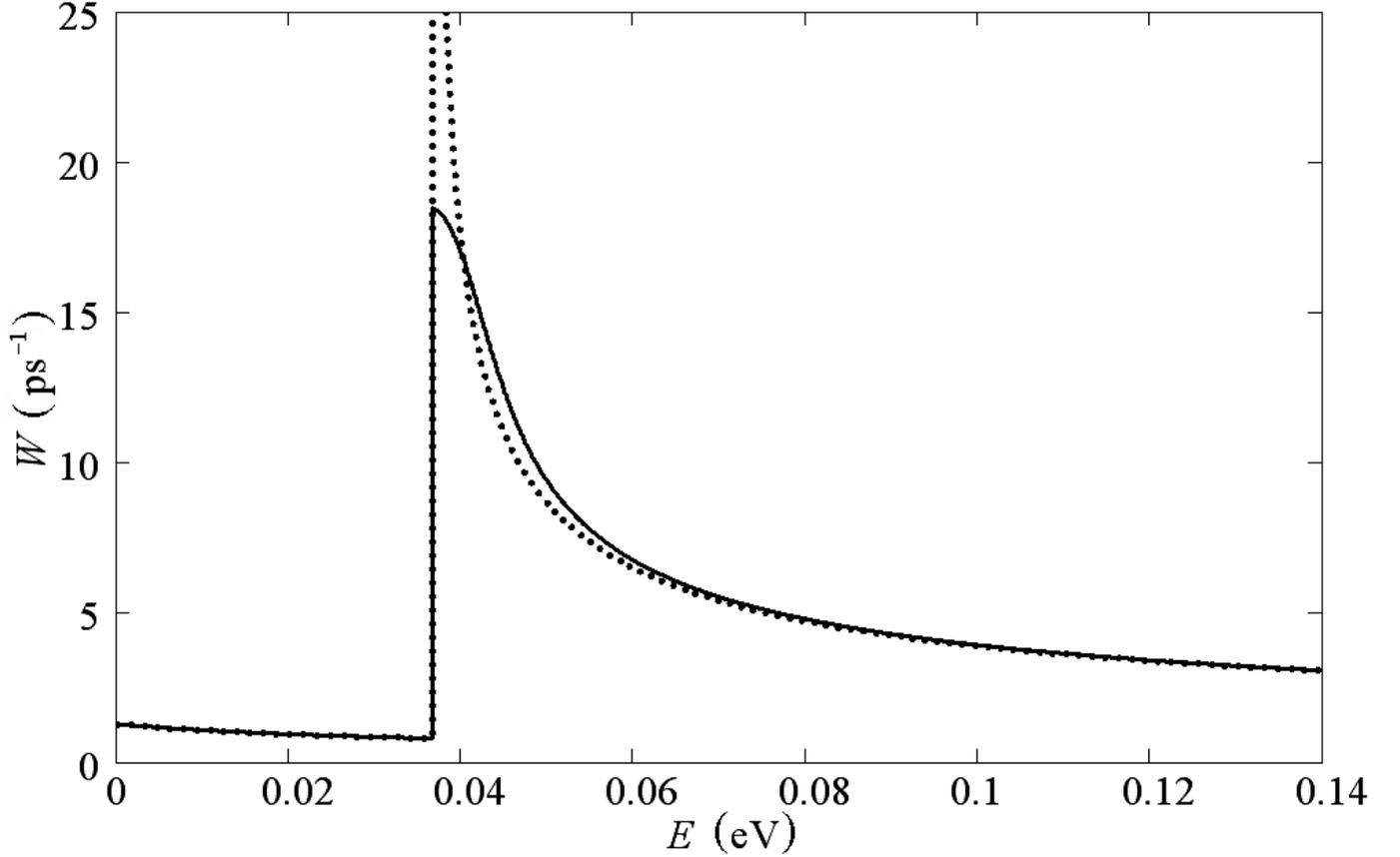

**Fig. 2** Polar optical phonon scattering rates of electrons in a GaAs quantum wire calculated under the conditions:
$T = 300$ K, $L_y = L_z = 100$ Å, $f(\pm k) = 0$.
*Dot line is the scattering rate calculated neglecting the collisional broadening.*
*Solid line is the scattering rate calculated with the collisional broadening taken into account.*

where $\hbar$ is the Plank constant; $k_B$ is the Boltzmann constant; $e$ is the elementary charge; $Z$ is the relative ion charge; $s$ is the sound velocity in GaAs; $\rho$ is the mass density of GaAs; $\varepsilon_s$ is the static dielectric permittivity of GaAs; $\varepsilon_\infty$ is the optic dielectric permittivity of GaAs; $\omega$ is the polar optical phonon frequency; $B_{ac}$ is the deformation potential of acoustic phonons; $m_x$ is the electron effective mass along the wire axis; $m_y$ and $m_z$ are the electron effective masses in directions O$y$ and O$z$, respectively; $T$ is the wire temperature; $L_y$ and $L_z$ are its transverse dimensions; $\delta$ is the root-mean-square deviation of the rough wire surface from the plane; $\Lambda$ is the roughness correlation length; $N$ is the ionized impurity linear concentration in quantum wire; $\eta$ is the non-parabolicity factor; $E_0$ is the energy level of the ground quantum state in GaAs; $E$ is the electron kinetic energy; $\Delta E$ is the collisional broadening factor (electron energy uncertainty [16, 17]); $W_1^{f/b}$ is the rate of forward/backward polar optical scattering with phonon absorption; $W_2^{f/b}$ is the rate of forward/backward polar optical scattering with phonon emission; $W_3^{f/b}$ is the rate of forward/backward acoustic scattering with phonon emission and absorption; $W_4^{f/b}$ is the rate of forward/backward surface roughness scattering; $W_5^{f/b}$ is the rate of ionized impurity forward/backward scattering; $\Theta$ is the unit step function; $f$ is the electron distribution function.

Formulae (1) – (12) allow the electron scattering rates to be calculated self-consistently taking into account the collisional broadening caused by scattering processes.

As an example, plotted in Fig.1 are the scattering rates in thin free standing GaAs quantum wire in the electric quantum limit as functions of electron kinetic energy. For the elastic (surface roughness and ionized impurity) and quasielastic (acoustic phonon) scattering mechanisms only the backward scattering is produced. This figure shows that there are no singularity points in the the scattering rates versus the electron energy.

Presented in Fig.2 is another result obtained for the conditions close to ones taken in Ref. [4] where only the polar optical phonon scattering is considered. The scattering rates have been calculated by using the Fock's approximation in the mentioned study. Comparison of our results with the results of Ref. [4] has evinced their close accordance. The slight discrepancies can be explained by the following reasons: in the present work (i) we take into account only the first modes of the one-dimensional phonons whereas the bulk phonons are considered in Ref. [4]; (ii) we assume that the electrons are confined by four infinitely high potential barriers in the quantum well whereas in Ref. [4] the electrons are confined by the electric field in one direction; (iii) the non-parabolicity is not taken into account in Ref. [4].

Thus in this paper the method of calculation of electron scattering rates in thin free standing GaAs quantum wires taking into account the collisional broadening is considered. Moreover, the obtained results are in good agreement with known theoretical insights [4, 5, 7, 10, 14–18]. In conclusion, we would like to note that our further purpose is to study the electric properties of the considered structures by means of the Monte Carlo simulation.